# Vectorial Imaging of the Photodissociation of 2-Bromobutane Oriented via Hexapolar State Selection

Masaaki Nakamura [1], Po-Yu Tsai[2], Shiun-Jr Yang[1], King-Chuen Lin[1,3,*], Toshio Kasai [1,4], Dock-Chil Che[5], Andrea Lombardi [6], Federico Palazzetti [6,*], and Vincenzo Aquilanti [6,7]

[1] *Department of Chemistry, National Taiwan University, Taipei 10617, Taiwan*

[2] *Department of Chemistry, National Chung-Hsing University, Taichung 402, Taiwan*

[3] *Institute of Atomic and Molecular Sciences, Academia Sinica, Taipei 10617, Taiwan*

[4] *Institute of Scientific and industrial Research, Osaka University, Ibaraki, Osaka 567-0047, Japan*

[5] *Department of Chemistry, Graduate School of Science, Osaka University, Toyonaka, Osaka 560-0043, Japan*

[6] *Università di Perugia, Dipartimento di Chimica, Biologia e Biotecnologie, 06123 Perugia, Italy*

[7] *Istituto di Struttura della Materia, Consiglio Nazionale delle Ricerche, 00016 Rome, Italy.*

*To whom correspondence should be addressed at kclin@ntu.edu.tw; federico.palazzetti@unipg.it

**Abstract**

Molecular orientation techniques are becoming available in the study of elementary chemical processes, in order to highlight those structural and dynamical properties that would be concealed by random rotational motions. Recently successful orientation was achieved for asymmetric-top and chiral molecules of much larger complexity than hitherto. In this work, we report and discuss the correlation between the vectors photofragment recoil velocity **v**, transition dipole moment **μ**, and permanent dipole moment **d** in a dissociation experiment on hexapole oriented 2-bromobutane, photoinitiated by a linearly polarized laser. The sliced ion images of the Br* ($^2P_{1/2}$) and Br ($^2P_{3/2}$) photofragment were acquired at 234.0 and 254.1 nm, respectively, by (2+1) resonance-enhanced multiphoton ionization technique. A detailed analysis of the sliced ion images obtained at a tilting angle 45º of the laser polarization provides the information on correlation of the three vectors, which are confined by two polar angles $α$, $χ$ and one azimuthal angle $φ_{μd}$ in the recoil frame. The sliced ion images of Br fragments eliminated individually from the enantiomers at 254.1 nm yield the asymmetric factor close to zero; for this reason the photofragment angular distributions do not show significant differences. The elimination of Br* fragment at 234.0 nm is mainly correlated with a parallel transition, giving rise to a large anisotropy parameter of 1.85, and thus can be considered as a single state excitation. The resulting recoil frame angles are optimized to 163±8° and 164±1° for $α$ and $χ$, respectively, whereas $φ_{μd}$ approaches close to 0º for the best fit. Since in the present case, the three vectors have an only slight spatial arrangement, the photofragment angular distributions of the two enantiomers do not show appreciable differences. Theoretical and computational simulations provide us the basis to state that oriented enantiomers can be discriminated on-the-fly in photodissociation processes even initiated by non-circularly polarized light, provided that three vectors encountered above have specific three-dimensional arrangements. The fact that Br fragment elimination involves a multi-potential dissociation carries uncertainties in theoretical estimates of vector direction. Therefore, this work represents a preliminary but significant step on the road of the chiral discrimination on-the-fly, which is shown to be best propitiated in molecules where vectors are far from having degenerate mutual angular directions.

# I. Introduction

Implementation of ion imaging technique in a gas-phase molecular photodissociation experiment permits analysis of scalar and vector properties of the phenomenon and the exploration of the complex dynamics involved. The employment of hexapolar electric field orientation technique induces preferential distribution of the permanent dipole moment of gas-phase molecules in specific directions and makes it possible to select the distribution of rotational states. The selection of a single rotational state has been achieved only for small and highly symmetric molecules, and has been applied to a series of diatomic and triatomic molecules, and to symmetric-top molecules[1–7]. An early example of orientation of an asymmetric-top molecule was recorded twenty years ago[8]. Relevant progress in orienting molecules of increasing complexity has been reported in the last decade.[9–11]

Recently, the asymmetric top molecule 2-bromobutane has been successfully oriented by the hexapole field technique[12, 13] and used to study the bromine atom photofragment distribution in photodissociation.[14] 2-bromobutane is a chiral molecule which was selected for the following reasons. First, photodissociation can take place in the wavelength region >230 nm. Second, the Br (or Br*) photofragment can be ionized via resonance-enhanced multiphoton ionization (REMPI) scheme followed by detection with ion-imaging technique. Third, the dissociation mechanism approaches axial recoil limit, for the Br (or Br*) elimination stemming from the repulsive states in the A band. Fourth, it is possible for manifestation of chirality of molecules while oriented by a hexapolar field. Traditional chiral discrimination methods reveal dichroism involving the distinct absorption cross section of enantiomers against circularly polarized lights: specifically, techniques of photoelectron circular dichroism detect molecular chirality by measuring photoelectron scattering distributions from synchrotron radiation (see for example Refs. 15-18 and references therein). By contrast, in this work we examine the possibility of enantiomeric discrimination on-the-fly by conducting photodissociation experiments on oriented chiral molecules using a linearly polarized laser. Even though the linearly polarized light is not a chiral light source, the combination with spatially-confined molecular orientation produces a chiral detection environment.[19-21]

Thus far, several methods have been successfully employed for the chiral discrimination such as circular dichroism, (cavity-based) optical rotary dispersion,[22] and (broad band or high-resolution) molecular rotational spectroscopy.[23] As is well known, both the target molecule and the probe radiation (e.g., circularly-polarized light) have to be chiral for differentiation of the molecular chirality in a conventional experiment. If either one is non-chiral, the experimental condition should be properly regulated to bring up an environment suitable for chiral discrimination. For instance, CO lacks the property of circular dichroism, but the CO molecules after orientation may possess circular dichroism in the measurement of photoelectron angular distribution.[24,25] In this sense, this work attempts a reverse way; i.e., the target molecule is chiral, but the light source is not and therefore chiral discrimination is possible in principle.

Indeed it has been predicted theoretically,[26,27] the experimental geometry of oriented molecules should provide an appropriate detection environment for chiral discrimination. Further, monitoring photofragment ion images offers the merits, including (1) the possibility for chirality recognition, and (2) molecular geometry confirmation by realizing the three-vector correlation in the recoil frame. Wörner and co-workers reported a similar work based on photodissociation to look into the chiral effect but using time-resolved photoelectron-circular dichroism (TR-PECD), which was performed on chiral CHBrFI and 2-iodobutane molecules, following selective C-I

bond dissociation.[28]

The photofragment angular distribution obtained from isotropic parent molecules essentially yield a single spatial anisotropy parameter $\beta$ with which the dissociation process can be gained insight:

$$I(\theta) = \frac{\sigma}{4\pi}[1 + \beta P_2(\cos\theta)] \qquad (1)$$

where $P_2$ is the second-order Legendre polynomials. $\beta$ is related to the transition dipole direction in the molecular frame within the axial recoil approximation, which implies that recoil times are so fast such that the asymptotic recoil velocity is directed along the initial bond axis for dissociation.[6,29–32] Photodissociation of oriented molecules monitored by the ion imaging technique can offer the information regarding three vector correlations, (1) photofragment recoil velocity **v**, (2) molecular transition dipole moment **μ**, and (3) the permanent dipole moment **d**. They are possible to be extracted and defined by two polar angles and one azimuthal angle in the recoil frame.[14,30,33] These three vector properties provide the possibility to discern enantiomers. Especially, the transition dipole moment **μ** is a transient property; therefore, the photofragment distribution would reflect the molecular chirality in a reaction. Moreover, those observations describe the dynamical figure of the chiral molecular photodissociation. Even for a non-axial recoil dissociation mechanism such as OCS,[6,7] the **v-μ-d** correlation can be used to shed light on the complicated pathways involved in multi-surface dissociation dynamics, because each potential surface owns particular topology and force field, thus probably contributing distinct features in vector correlations.

In this work, photodissociation dynamics of 2-bromobutane enantiomers in S- and R- form, each oriented via a hexapole state selector, is investigated, and we attempt the possibility to distinguish these two enantiomers using linearly polarized laser. Under the condition that the laser polarization was tilted at 45° towards the detector surface, the slicedion images of the Br* ($^2P_{1/2}$) and Br ($^2P_{3/2}$) photofragment were acquired at 234.0 and 254.1 nm, respectively, by (2+1) REMPI technique. Then, the sliced ion images are analyzed to obtain how the above-mentioned three vectors are correlated in the recoil frame. They are defined by two polar angles $α, χ$ and one azimuthal angle $φ_{μd}$ in the recoil frame. Finally, we will discuss the conditions for chiral discrimination.

## II. Experiment

In this sliced ion imaging setup, the 2-bromobutane at about 65 Torr at room temperature was expanded through a pulsed valve. The skimmed molecular beam was then focused and rotationally selected by a 70-cm long hexapole state selector which was biased up to 4.0 kV. The transmitted molecules were in turn oriented by means of an electrostatic field adjusted at about 200 V/cm; this field was also used as the ion extraction field in a two-stage ion lens such that the orientation direction is along the time-of-flight (TOF) axis of the molecular beam. The oriented molecules were photolyzed with a linearly polarized laser and the resulting Br or Br* photofragments were successively ionized by a (2+1) REMPI scheme with the same laser. The ion clouds were velocity-mapped onto a two-stage microchannel plate (MCP) coupled with a phosphor screen, followed by the emission detection using a charge coupled device camera. For further details, see Ref. 14.

The present experiments were conducted with two 2-bromobutane enantiomers: R- and S-form of 85% and 92% purities, respectively. These enantiomeric samples were synthesized by taking corresponding 2-butanol enantiomer as the precursor and the individual purity was evaluated by using gas chromatography with the capillary column for enantiomer separation (Macherey-Nagel LIPODEX® C).

A 308 nm excimer laser-pumped dye laser, operated synchronously with the pulsed valve at 10 Hz, was frequency-doubled to emit UV light. The (2+1) REMPI ion signals of Br*($^2P_{1/2}$) and Br($^2P_{3/2}$) were acquired at 234.0 and 254.1 nm, respectively. The former laser energy was adjusted to 200 μJ/pulse, while the latter one was 500 μJ/pulse. The laser polarization was adjusted at a tilting angle of 45° with respect to the time-of-flight axis. The resulting sliced ion images of Br* or Br photofragment were acquired in photolysis of parent molecules oriented by the hexapole field. The MCP detector was gated with a 25 ns-pulsed voltage to observe only the central slice of the $^{79}$Br ion cloud. Although the gate pulse is positioned to slice the $^{79}$Br isotope, trace of $^{81}$Br still emerges. While taking a series of sliced ion images of both isotopes by adjusting the delay time of the pulsed gate, as shown in Fig.1, it turns out that the later half-sphere of the $^{79}$Br ion cloud overlaps with the earlier half-sphere of $^{81}$Br, thereby confirming that the central region of the image is due to the presence of $^{81}$Br.

## III. Results and Discussion

Preliminary to the presentation of the photodissociation experiments, an analysis of the contributing conformers is in order: in fact, the focusing curve of the 2-bromobutane molecular beam is mainly contributed by three conformers, T, G+ and G-,[14,34] whose sum in the simulation is consistent with the experimental result. The simulated focusing curve takes into account rotational states up to 22 at the upper limit of 5.0 kV. Their rotational structures are congested for a molecule of such a complexity, making the selection of a single rotational state out of reach and uninformative. However, the hexapole state selector reveals a non-statistical rotational distribution as a signature of molecular orientation, a key feature to be exploited for exploring the complex photodissociation dynamics of molecules where the symmetric-top symmetry is broken.

## A. Sliced ion imaging of photofragments of oriented molecules at 234 nm

Under the condition of random orientation for 2-bromobutane molecular beam, sliced ion images of Br* have been acquired at 234.0 nm and the corresponding anisotropy parameter $\beta$ was is determined to be 1.85±0.07.[14] There are three major electronic states contained in the A band leading to the C-Br bond dissociation following the photolysis. The Br* fragment was obtained via a parallel transition, thereby yielding the positive anisotropy parameter which may approach to the limit of 2. Thus, a single excited state involvement is considered to dominate the photodissociation process.[33]

When the laser polarization is rotated 45° toward the MCP detector surface, the oriented 2-bromobutane in either R- or S-form of enantiomer is photolyzed and the resulting sliced ion images of the Br* photofragments are shown in Fig.2(a). Expansion in Legendre polynomials describes the angular dependence of the experimental results:[6,7] the leading terms are

$$I(\theta) = 1 + \beta_1 P_1(\cos\theta) + \beta_2 P_2(\cos\theta) \qquad (2)$$

where $P_1$ and $P_2$ are the first- and the second-order of Legendre polynomials, while $\beta_1$ and $\beta_2$ indicate asymmetric factor and anisotropy parameter, respectively. The values of $\beta_1$ and $\beta_2$ are measured as 0.24±0.05 and 1.59±0.02, respectively. If the linear polarization of the incident laser beam is turned to 90° with respect to the time-of-flight axis, the $\beta_1$ factor disappears, independently on any bias from molecular orientation. In this case, Eq. (2) is simplified to Eq. (1) and $\beta_2$ is equal to $\beta$. When the experimental conditions change, the $\beta_2$ value becomes smaller than $\beta$, but cannot be the same as $\beta$. $\beta_1$ and $\beta_2$ are associated with the angles $\alpha$ and $\chi$, where $\alpha$ is the angle between the permanent dipole moment **d** and the fragment recoil velocity **v**, while $\chi$ is the angle between the transition dipole moment **μ** and **v**. These coefficients are determined from the following equation, which simplifies when the laser polarization is tilted to 45º,[7,30]

$$\beta_1 = \frac{3c_1 \sin\alpha \sin\chi \cos\chi}{[1+2c_2 P_2(\cos\alpha)][1-(1/2)P_2(\cos\chi)]+(3/16)c_2 \sin^2\chi \sin^2\alpha} \quad (3)$$

$$\beta_2 = \frac{[1+2c_2 P_2(\cos\alpha)]P_2(\cos\chi)+(3/8)c_2 \sin^2\chi \sin^2\alpha}{[1+2c_2 P_2(\cos\alpha)][1-(1/2)P_2(\cos\chi)]+(3/16)c_2 \sin^2\chi \sin^2\alpha} \quad (4)$$

where $c_1$ and $c_2$ are the first and the second moment of the Legendre expansion in the expression of the orientation distribution. The parameters $c_1$ and $c_2$ were determined previously from the time-of-flight measurements.[12,27] $\beta_1$ and $\beta_2$ are measurable, while $\alpha$ and $\chi$ are to be determined through (3) and (4). Therefore, we are able to obtain the recoil frame angles and consequently disentangle a complex photodissociation dynamics, given the experimental results. Note that Eqs. (3) and (4) are based on the assumption that three vectors **v**, **d** and **μ** are coplanar; the analysis of the general case, most relevant in the following, will be discussed in the next section. Additionally, if more than one transition is involved in the photodissociation, further consideration is required and will be discussed later.

### B. Recoil frame angle analysis

The sliced ion images are associated with the vector correlation of photofragments in photodissociation. Hence, the recoil frame angles can be obtained from the analysis of experimental results. The recoil frame is composed of three vectors representing the permanent electric dipole **d** for a polyatomic molecule in its electronic ground state, the recoil velocity **v** of the photofragments and the transition dipole moment **μ** in a photodissociation process induced by single state excitation with linearly polarized light. The three vectors **v**, **d** and **μ** are defined in the molecular recoil frame xyz, where **v** lies along the z-axis and **d** is in the xz-plane. Vectors **v** and **d** define the polar angle $\alpha$, and the vectors **v** and **μ** define the other polar angle $\chi$, and $\varphi_{\mu d}$ is the azimuthal angle defined by the projection of **μ** onto the xy-plane and the x-axis. The $\varphi_{\mu d}$ should be 0 for linear top or symmetric top molecules. For chiral molecules, the sign of $\varphi_{\mu d}$ may be either positive or negative for a specific enantiomer, because of a potential symmetry breaking. [14,30]

The sign of $\varphi_{\mu d}$ in the recoil frame can be probably observed in the photofragment angular distribution of oriented chiral molecules as photolyzed by linearly polarized light. The theoretical formulation of the angular distribution in terms of these recoil frame angles[30]:

$$I(\gamma, \delta, \varphi_{O\varepsilon}) = [1 + 2P_2(\cos\chi)P_2(\cos\gamma)][1 + 2c_1(\cos\alpha)(\cos\delta)$$
$$+ 2c_2 P_2(\cos\alpha)P_2(\cos\delta)]$$
$$+ 6\sin\gamma\cos\gamma\sin\chi\cos\chi\sin\alpha\sin\delta\cos(\varphi_{O\varepsilon} + \varphi_{\mu d})[c_1$$
$$+ 3c_2\cos\alpha\cos\delta] + \frac{9}{8}c_2\sin^2\gamma\sin^2\chi\sin^2\alpha\sin^2\delta\cos 2(\varphi_{O\varepsilon} + \varphi_{\mu d})$$

(5)

where $\gamma$, $\delta$, and $\varphi_{O\varepsilon}$ are the other recoil frame angles (see Fig.3), $P_2$ is the second order Legendre polynomial, and $c_1$, $c_2$ are the expansion coefficients of the molecular orientational probability distribution, determined previously to be -0.35 and 0.5, respectively.[14,33] For understanding the angle parameters expressed in Eq. (5), a recoil frame is shown in Fig. 3, similar to the one reported previously.[14] For simplicity, the terms up to the second order were taken into account. The negative sign of $c_1$ does not appear in Ref.14; it arises to account for the difference in the definition of laboratory frame geometry described below, and thereby it does not make any essential difference.

The photofragment distribution in the recoil frame can be transformed to the expression in the laboratory frame by substituting the following equations into Eq. (5).[30]

$$\cos \gamma = \cos \Omega \cos \Gamma + \sin \Omega \sin \Gamma \cos \Theta \qquad (6)$$

$$\cos \delta = \cos \Omega \cos \Delta + \sin \Omega \sin \Delta \cos(\Theta - \Phi) \qquad (7)$$

$$\cos \varphi_{O\varepsilon} = \{\sin^2 \Omega \cos \Gamma \cos \Delta + \sin \Gamma \sin \Delta \cos \Phi - \sin \Omega \cos \Omega [\sin \Delta \cos \Gamma \cos(\Phi - \Theta) + \sin \Gamma \cos \Delta \cos \Theta] - \sin^2 \Omega \sin \Gamma \sin \Delta \cos(\Phi - \Theta) \cos \Theta\}/(\sin \gamma \sin \delta) \qquad (8)$$

$$\sin \varphi_{O\varepsilon} = \{\cos \Omega \sin \Gamma \sin \Delta \sin \Phi - \sin \Omega [\sin \Delta \cos \Gamma \sin(\Phi - \Theta) + \sin \Gamma \cos \Delta \sin \Theta]\}/(\sin \gamma \sin \delta) \qquad (9)$$

Here, $\Gamma$ is the tilt angle of the photolysis laser polarization, $\Delta$ and $\Phi$ the elevation and azimuth angle of the orientation field, respectively, as well as $\Omega$ and $\Theta$ the elevation and azimuth angle of the recoil velocity. All these angle parameters are with respect to the time-of-flight (TOF) axis. The recoil frame angles in Eq. (5), $\gamma$, $\delta$ and $\varphi_{O\varepsilon}$, are replaced with the laboratory frame angles. Accordingly, the sliced ion imaging detection with the current apparatus in the laboratory coordinates, shown in Fig. 4, is defined as follows; the TOF axis coincides with the Z-axis, the laser propagates along the Y-axis to intersect the oriented molecular beam inside the ion lens, and the recoil velocities of fragments for detection lie in the XY-plane ($\Omega = 90°$). The polarization direction of the laser $\varepsilon$ is tilted at 45° with respect to the Z-axis in the XZ-plane ($\Gamma = 45°$). The orienting electric field **O** is identical to the ion extraction field; thus it is parallel to the TOF axis and points toward the detector ($\Delta = 0°$, $\Phi = 0°$). The molecules are oriented in the orienting field by its permanent dipole moment before the photodissociation. The sign of $c_1$ represents the direction of molecular orientation with respect to the orienting field. The orienting field vector **O** points from the high voltage side to the low voltage side of the electric field, and the molecular permanent dipole moment **d** points from the negative to the positive charged distribution of the molecule.

$c_1$ is positive when **O** and **d** vectors are directed toward the same way, and *vice versa*.

As a result, the photofragment angular distribution of oriented chiral molecules obtained with sliced ion imaging technique can be expressed as an expansion of associated Legendre polynomials,

$$I(\Theta) = 1 + \beta_1^0 P_1^0(\cos\Theta) + \beta_2^0 P_2^0(\cos\Theta) + \beta_1^1 P_1^1(\cos\Theta) + \beta_2^1 P_2^1(\cos\Theta)$$

(10)

where $P_l^m$ is the associated Legendre polynomials, $\Theta$ is the azimuthal angle of the velocity vector **v** onto a plane parallel to the XY-plane, and the coefficients $\beta_l^m$ are determined by $b_l^m/b_0^0$:

$$b_0^0 = \left(1 - c_2 P_2(\cos\alpha)\right)\left(1 - \frac{1}{2}P_2(\cos\chi)\right) + \frac{3}{16}c_2 \sin^2\alpha \sin^2\chi \cos 2\varphi_{\mu d}$$

(11)

$$b_1^0 = 3c_1 \sin\alpha \sin\chi \cos\chi \cos\varphi_{\mu d}$$ (12)

$$b_2^0 = \left(1 - c_2 P_2(\cos\alpha)\right)P_2(\cos\chi) + \frac{3}{8}c_2 \sin^2\alpha \sin^2\chi \cos 2\varphi_{\mu d}$$

(13)

$$b_1^1 = -\frac{9}{8}c_2 \sin^2\alpha \sin^2\chi \sin 2\varphi_{\mu d}$$ (14)

$$b_2^1 = -c_1 \sin\alpha \sin\chi \cos\chi \sin\varphi_{\mu d}$$ (15)

For an ideal sliced ion imaging experiment, the recoil velocity vectors of observed photofragments lie in the XY-plane. Therefore, $\Theta$ directly corresponds to the angle from the vertical axis on a sliced image. Thus, the experimental observation can be fitted with Eq. (10).

The series of associated Legendre polynomials $P_l^m(\cos\Theta)$ are characterized by two integers, *m* and *l*, where *l* denotes the degree of the polynomials and both *l* and *m* are related to the nodes of the functions. Specifically, the *m*=0 case is that of the Legendre polynomial $P_l(\cos\Theta)$. $P_l^m(\cos\Theta)$ is an odd function of $\Theta$ when $m$ is odd, while it is an even function when $m$ is even. All the polynomials $(P_1^0, P_2^0, P_1^1, P_2^1)$ appearing in Eq. (10) have different symmetries in the range of $0° \leq \Theta < 360°$, four quantities $(\beta_1^0, \beta_2^0, \beta_1^1, \beta_2^1)$ can be obtained from experimental observations, and thus the three recoil frame angles ($\alpha$, $\chi$, $\varphi_{\mu d}$) can be determined accordingly.

The last two terms of Eq. (10), $\beta_1^1$ and $\beta_2^1$ are non-zero only for chiral molecules, with different sign for each enantiomer, being the sign of $\varphi_{\mu d}$ either positive or negative depending on its handedness, as discussed above. Eq. (10) is an even function without the last two terms, in consequence, the corresponding sliced image would be axis-symmetric to the vertical axis, as in the case of ordinary photofragment image of non-oriented molecules. The molecular chirality breaks the symmetry to introduce the contributions of those odd functions; $P_1^1$ and $P_2^1$. Fig.5(a) shows a schematic recoil frame of enantiomers. The recoil frames **d-v-μ** and **d-v-μ'** are confined by the angles $\alpha$, $\chi$ and $\varphi_{\mu d}$ with the same magnitude; however, the sign of $\varphi_{\mu d}$ is different. As a result, the two frames are distinguishable. Given an example with $\alpha = 45°, \chi = 45°,$ and $\varphi_{\mu d} = \pm 150°$, the two photofragment images are simulated in Fig.5(d), and the corresponding angular distributions are displayed in Fig.5(e), which give rise to significant angular shift between these two enantiomers. The images are no longer axis-symmetric, and the maxima of the angular distributions are shifted from the top, $\Theta = 0$, to opposite sides. The angular shifts verify chirality of molecules and its role in a photodissociation reaction. If the parent molecule is achiral or non-oriented, this kind of symmetry breaking may not show up. Also, if the sample is racemic, the equal mixture of enantiomers, the symmetry breaking may be canceled out each other.

It is important to note that not always asymmetric-top molecules can be distinguished from these photofragment distributions. For example, if $\varphi_{\mu d}$ equals to either 0° or 180°, $\beta_1^1$ and $\beta_2^1$ are zero, and the recoil frames **d-v-μ** are coplanar as shown in Fig.5(b). In this case, the two mirrored recoil frames **d-v-μ** and **d-v-μ'** cannot be distinguished and thus angular distributions are symmetric around $\Theta = 0$. Another case is shown in Fig.5(c). If either $\alpha$ or $\chi$ are close to 0°, the geometry of three vectors approximates to coplanar even if $\varphi_{\mu d}$ is far from 0° or 180°. This kind of geometry also makes it difficult to reveal molecular chirality effects, because of unavoidable restriction of the sensitivity of the angular distribution. In both cases, the angle $\varphi_{\mu d}$ nearly loses its correlation to the photofragment distribution; therefore, the determination of $\varphi_{\mu d}$ may turn out increasingly difficult. However, even if the last two terms in Eq. (10) cancel out, the recoil frame angles, $\alpha$ and $\chi$, can be evaluated by the other terms.

In this work, the ion image of Br* at 234.0 nm was exemplified for analysis. Because of its large $\beta$ value (1.85),[14] the photodissociation may be considered to be caused by a single parallel transition, and the nonadiabatic contribution is negligible. As shown in Fig.2(a), the sliced ion images of Br* of 2-bromobutane in either S- or R-form of the enantiomers were acquired individually at 234.0 nm, but their corresponding angular distributions in Fig.2(c) as a function of $\Theta$ fail to show any significant difference. As described in the above vector model, the result implies that $\varphi_{\mu d}$ is extremely small ($\varphi_{\mu d} \sim 0$) or either $\alpha$ or $\chi$ angle were close to 0° or 180° such that the different sign of $\varphi_{\mu d}$ cannot be separated. The sliced ion image at 234.0 nm presents positive $\beta_1$; if either $\alpha$ or $\chi$ was close to 0°, $\beta_1$ should also be close to zero. Thus the small $\varphi_{\mu d}$ will be reasonable to explain the result. Eq. (10) is then simplified to Eq. (2), retaining only the first three $P_l^m$ terms which are related with the $\alpha$ and $\chi$ angles. Given the anisotropy coefficients $\beta_1$ and $\beta_2$ and the orientation coefficients $c_1$ and $c_2$ obtained previously,[12] $\alpha$ and $\chi$ are optimized to 163±8° and 164±1°, respectively. The obtained $\chi$ is very close to the previous result χ = 167°, but $\alpha$ is deviated from 139° reported.[14] The deviation may arise from the different experimental conditions. In this work, sliced ion images with a narrow gate, 25 ns are acquired; this is in contrast to the previously used wider gate about 400 ns adopted to cover the whole isotopic variant of bromine. The difference results in a slightly deviated spread of the kinetic energy release. The previous result was obtained by taking into account a thin shell of the outermost ring from a fully-projected image which is approximated to a center slice of $^{79}$Br* fragment. However, such a treatment might be affected by off-center contributions because of the velocity spread. The off-center contribution generally intensifies the $\beta_1$ value to some extent, and thereby $\alpha$ tends to be closer to 90°.

As with the photolysis at 234.0 nm, the sliced ion images of Br fragments originated from enantiomers in S- or R-form were individually acquired and their corresponding angular distributions as a function of Θ appear to be similar, so that no significant angular discrimination is found. The three vectors are expected to lie on coplanar, and $\varphi_{\mu d}$ approaches 0°. The Fig. 2(b) shows the simulated sliced ion image with the obtained recoil frame angles. It shows a good agreement with the experimental image in Fig. 2(a).

C.  **Sliced ion imaging of photodissociation of oriented molecules at 254.1 nm**

A sliced ion image of the ground state $^{79}$Br photofragment was acquired by photolysis of non-oriented 2-bromobutane at 254.1 nm. As with the Br* image obtained at 234.0 nm, a fraction of the other isotope $^{81}$Br, which could not be fully separated, may contribute to the central region. The analysis of the angular distribution yields an anisotropy parameter $\beta$ of 0.93±0.03, which is much smaller than the value of 1.85 obtained at 234.0 nm, showing a mixing of perpendicular- and parallel-transition characters.[33] The ground state Br is correlated with the $^3Q_1$ state via a perpendicular transition. When the photolysis wavelength is longer, this transition tends to be favored. The $^3Q_0$ excitation, which leads to the Br* fragment via a parallel transition, is found to have a much larger absorption cross section for the feature of the A band. As a result, the $\beta$ value for the Br fragment is characteristic of a mixed transition containing perpendicular excitation mainly ascribed to the $^3Q_1$ state and a nonadiabatic transition contributed by the coupling between $^1Q_1$ and $^3Q_0$.[35,36]

Since the observed anisotropy parameter $\beta$ is caused by a mixture of perpendicular and parallel transitions, the mixing ratio of these two components can be estimated by,

$$\beta_{\text{obs}} = r\beta_\parallel + (1-r)\beta_\perp \qquad (16)$$

where $r$ is the fraction of the parallel-transition component. Its value is obtained to be 0.64, given $\beta_\parallel = 2$ and $\beta_\perp = -1$ at their extreme limit.

When the 2-bromobutane molecule was oriented by hexapole field at 4.0 kV, the sliced-ion image of the Br photofragment was acquired in photolysis at 254.1 nm, as shown in Fig.6(a). Analysis of the angular distribution for the sliced image gives rise to the best fit for $\beta_1$ and $\beta_2$ to be 0.00±0.02 and 0.57±0.02, respectively. As with the above-mentioned $\beta$ value, a small $\beta_2$ results from the contribution of at least two transitions. One bears the parallel-transition character contributed by the nonadiabatic coupling between the $^1Q_1$ and $^3Q_0$ state, while the other is mainly contributed by the $^3Q_1$ state with perpendicular-transition character.

Two recoil frame angles $\alpha$ and $\chi$ are required in the simulation for one transition. Since the elimination of the Br fragment involves the contribution of both parallel and perpendicular transitions, it is necessary to take into account two sets of the angles $\alpha$ and $\chi$, ($\alpha_\parallel$, $\chi_\parallel$) and ($\alpha_\perp$, $\chi_\perp$). Upon the photolysis, the C-Br bond fission associated with the A band transition may well fit the axial recoiling condition, and thus $\alpha_\parallel$ becomes identical to $α_⊥$, because the permanent dipole moment is a static property of the molecular structure. The $α = 163°$ is used, consistent with that at 234.0 nm. As for $χ$, it is reasonable to consider $χ_∥ = 0°$ for the parallel-transition component and $χ_⊥ = 90°$ for the perpendicular-transition component.

When the laser polarization was tilted at 45° as described above, given the mixing ratio $r = 0.64$, Fig.6(b) shows the simulated counterpart of the sliced ion image of Br obtained at 254.1 nm by integrating the simulation individually for both parallel-transition $I_∥(α, χ_∥ = 0°)$ and perpendicular-transition $I_⊥(α, χ_⊥ = 90°)$ components. The corresponding fragment angular distribution yields $β_1 = 0.00$ and $β_2 = 0.43$, which appear to be satisfactorily consistent with the values of $β_1 = 0.00±0.02$ and $β_2 = 0.57±0.02$ obtained from the best fit of the experimental data shown in Fig.6(c) based on Eq. (2). Especially, $β_1$ is almost zero, indicating the lack of asymmetry. That is why the Br fragment angular distributions between the enantiomers in R- and S-form turn out to be not distinguishable.

## IV. Summary and Final Remarks.

Sliced ion images of Br* ($^2P_{1/2}$) and Br ($^2P_{3/2}$) have been acquired at 234.0 and 254.1 nm respectively, in the photodissociation of 2-bromobutane molecules oriented via a hexapole state selector coupled with a homogenous electric field. Three-vector correlation in the recoil frame is extracted, consisting of the photofragment recoil velocity **v**, molecular transition dipole moment **μ**, and permanent dipole moment **d**. They define two polar angles $α$, $χ$ and one azimuthal angle $φ_{μd}$ (Fig. 3). The elimination of Br* fragment is mainly correlated with a parallel transition with a large anisotropy parameter, and thus can be treated as a single state excitation. The optimal recoil frame angles $α$, $χ$ and $φ_{μd}$ are evaluated, thus shedding light on the geometrical features of molecules and spatial aspects of molecular dynamics when two of the three vectors **v**, **μ**, and **d** tend to be nearly collinear (corresponding to a nearly planar configuration), harder becoming the possibility of discrimination of the two enantiomers on-the-fly. In the present case, the sliced ion images of Br fragments eliminated individually from the two enantiomers yield corresponding angular distributions for on-the-fly discrimination undiscernible experimentally. This is additionally a multi-potential photodissociation process involving perpendicular transitions and nonadiabatic contributions so that the resulting recoil frame angles cannot be evaluated accurately unless an additional sliced ion imaging detection is conducted at a different rotation angle.

We believe that this paper presents an extensive assessment of experimental requirements for achieving chiral discrimination on-the-fly, that must be performed on molecules where vectors are arranged far from mutually degenerate orientation displacement in order to enhance the differences in angular distributions of the photofragments produced by the two enantiomers.

## V. Acknowledgments

This work is supported by the Ministry of Science and Technology of Taiwan, Republic of China under Contract No. NSC 102-2113-M-002009-MY3. T. K. thanks National Taiwan University for providing him a visiting professorship to carry out this work. The Japanese Ministry of Education, Science, and Culture is gratefully acknowledged for a Grant Aid for Scientific Research (No. 17KT0008) in support of this work. F. P., A. L. and V. A. acknowledge the Italian Ministry for Education, University and Research, MIUR, for financial supporting: SIR 2014 "Scientific Independence for young Researchers" (RBSI14U3VF).

**Figure Captions**

**Fig.1** The sliced ion images of Br* photofragments with different gate time delays. The photolysis laser polarization was set parallel to the TOF axis. The reported delay time shifts are measured from the center of the $^{79}$Br sphere.

**Fig.2** The sliced ion images of Br* photofragment from oriented 2-bromobutane enantiomers dissociated at 234.0 nm

**Fig.3** The geometry in the recoil frame. The recoil velocity **v** corresponds to the *z*-axis. The recoil velocity **v** and the laser polarization **ε** define the *zx*-plane. Couples of angles given in parentheses denote spherical polar angles of each vector: γ is the angle between **v** and **ε**, δ is that between **v** and the direction of the orienting field **O**; $\varphi_{O\epsilon}$ is defined between **O** and **ε**; finally, $\varphi_{\mu d}$ is the angle included between **μ** and **d**.

**Fig.4** The geometry of the experimental setup in the laboratory frame. The angle Θ is an azimuthal angle projected onto the XY-plane. For an ideal sliced ion imaging experiment, the observed recoil velocity vectors are in the XY-plane. The +X direction points to the vertical top. The laser propagates to +Y direction. The time-of-flight axis is parallel to the Z-axis and photofragments are accelerated to +Z direction. The laser polarization ε is tilted 45° from the X-axis in the XZ-plane. The orienting field vector is parallel to the Z-axis.

**Fig.5** Cases (a) (b) (c) show various geometry of recoil frame vector properties, **v**, **d**

and **μ**. (a) is a case of distinguishable **d-v-μ** and **d-v-μ'** frames. The three vectors point along different directions in space. (b) is an indistinguishable case, since **d-v-μ(μ')** are coplanar. (c) is also a case that is difficult to disentangle. $\varphi_{\mu d}$ is neither 0° nor 180°, but vectors are almost coplanar being $\alpha$ or $\chi$ ($\alpha$ in the upper figure) very small. (d) (e) The simulated images and angular distributions of (a) and (b). A coplanar **d-v-μ** geometry (b) has its maximum at 0°. However, the maxima of angular distributions of **d-v-μ** and **d-v-μ'** in (a) shift conversely along the angle axis. ($\alpha = 45°, \chi = 45°, \varphi_{\mu d} = \pm 150°$)

**Fig.6** The ion sliced ion images of Br photofragment from oriented 2-bromobutane enantiomers dissociated at 254.1 nm.

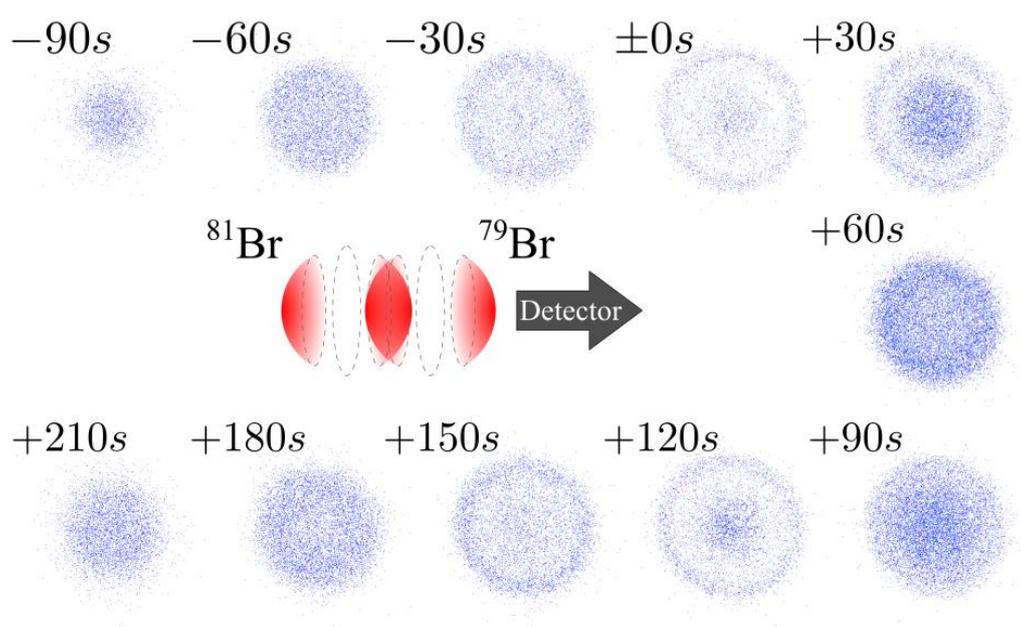

**Fig.1**

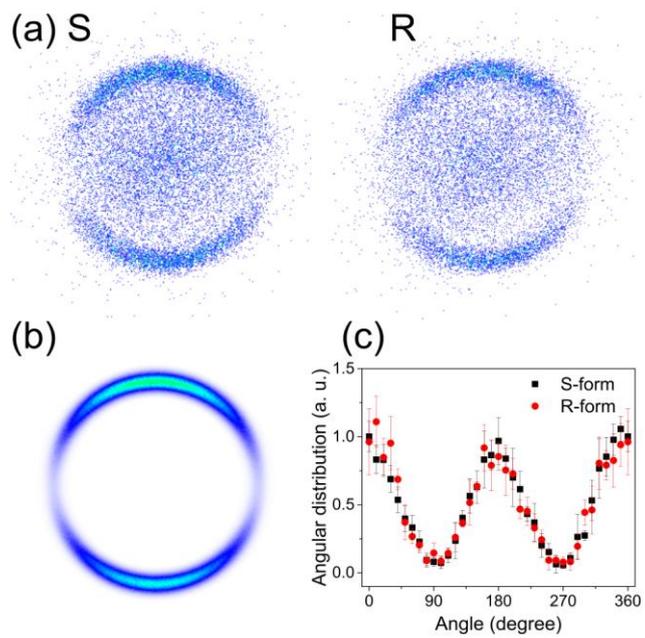

**Fig.2**

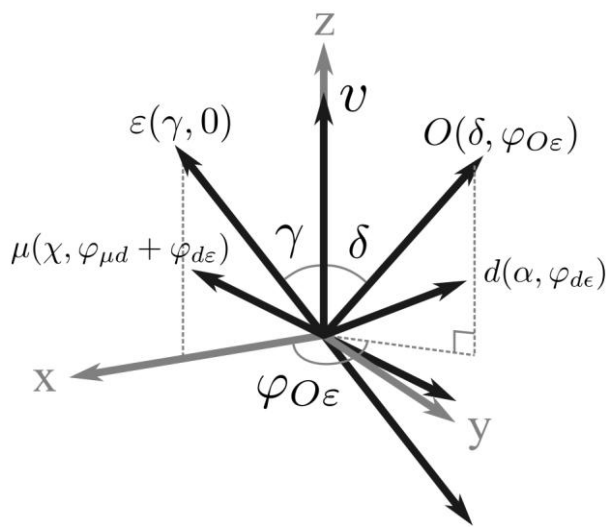

**Fig.3**

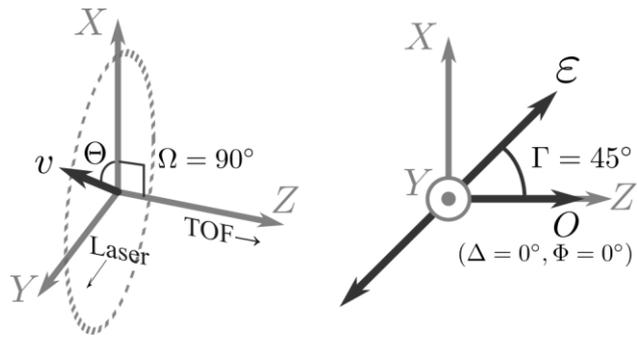

**Fig.4**

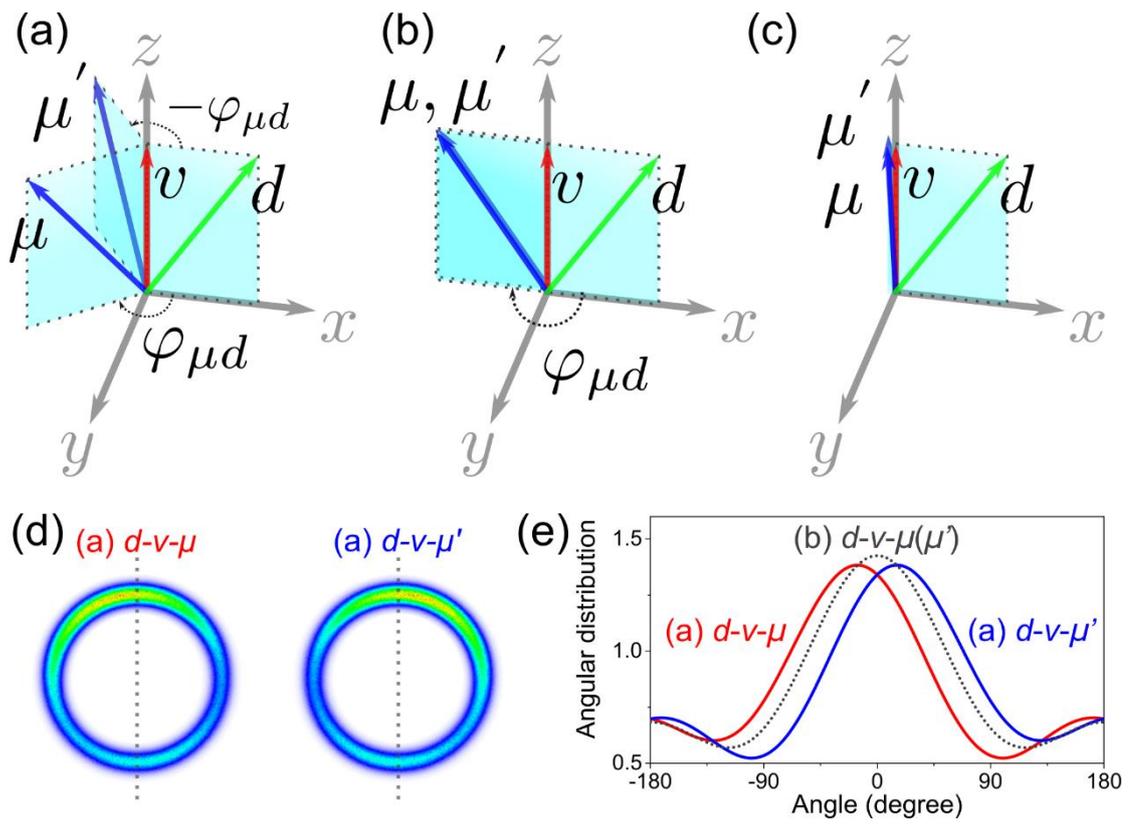

Fig.5

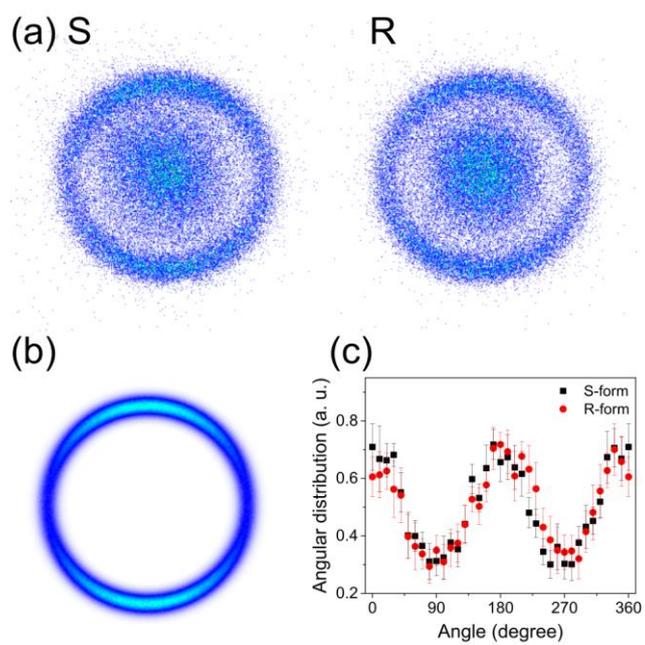

**Fig.6**